\documentclass[twocolumn,showpacs,preprintnumbers,amsmath,amssymb]{revtex4}

\usepackage{graphicx}% Include figure files
\usepackage{dcolumn}% Align table columns on decimal point
\usepackage{bm}% bold math
\usepackage{times}

\begin{document}

\title{Charge-Density-Wave Ordering in the Metal-Insulator Transition Compound PrRu$_4$P$_{12}$}

\author{C. H. Lee}
\author{H. Matsuhata}
\author{H. Yamaguchi}
\affiliation{National Institute of Advanced Industrial Science
and Technology, 1-1-1 Umezono, Tsukuba, Ibaraki 305-8568, Japan}
\author{C. Sekine}
\author{K. Kihou}
\author{T. Suzuki}
\author{T. Noro}
\author{I. Shirotani}
\affiliation{Muroran Institute of Technology, 27-1 Mizumoto, Muroran
050-8585, Japan}
\date{\today}

\begin{abstract}
X-ray and electron diffraction measurements on the metal-insulator (M-I) transition compound PrRu$_4$P$_{12}$ have revealed the emergence of a periodic ordering of charge density around the Pr atoms.  It is found that the ordering is associated with the onset of a low temperature insulator phase.  These conclusions are supported by the facts that the space group of the crystal structure transforms from Im$\bar{3}$ to Pm$\bar{3}$ below the M-I transition temperature and also that the temperature dependence of the superlattice peaks in the insulator phase follows the squared BCS function.  The M-I transition could be originated from the perfect nesting of the Fermi surface and/or the instability of the $f$ electrons.
\end{abstract}


\maketitle

Filled skutterudite compounds RM$_4$X$_{12}$ (R = rare-earth, M = Fe, Ru or Os; X = P, As or Sb) have attracted a great deal of interest in view of the origin of their dramatically variable physical properties.  For example, they show metal-insulator (M-I) transitions \cite{sekine97}, superconductivity \cite{shirotani97,meisner,bauer} and large thermoelectric performance \cite{sales} depending on the combination of elements.  To seek out the origin of this anomalous behavior, one of the important factors that must be understood is the topology of the Fermi surface.  According to de Haas-van Alphen measurements, the shape of the Fermi surface in LaFe$_4$P$_{12}$ is nearly cubic \cite{sugawara}.  This implies the presence of nesting instability with a wave vector of $\bf{q}$ = (1,0,0).  It was, thus, conjectured that some of the exotic properties in filled skutterudites originate from the nesting of the Fermi surface.  To confirm this hypothesis, an intensive effort has been put.  However, no clear evidence of nesting-induced phenomena has been identified so far.
%=========================================================
\begin{figure}
\includegraphics[width=\columnwidth]{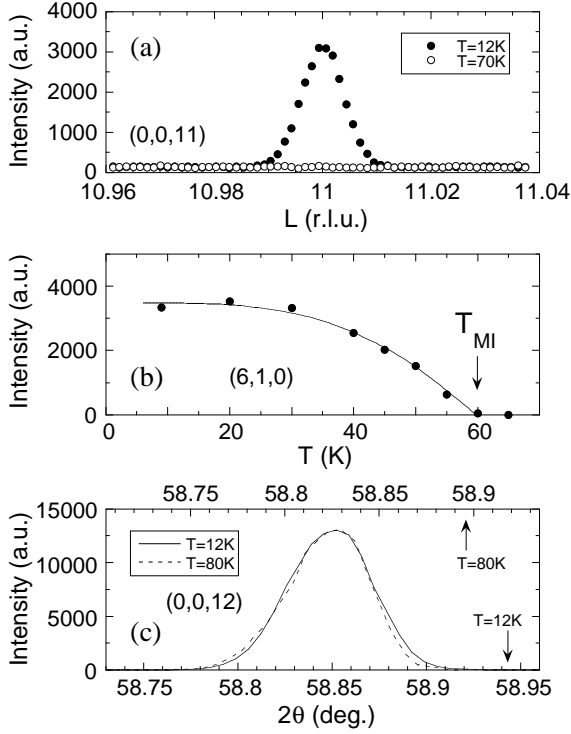}
\caption{\label{fig:satelite} (a) Longitudinal scans of a superlattice peak at (0,0,11) above and below $T_{MI}$ for single crystals of PrRu$_4$P$_{12}$ measured by x-ray diffraction.  (b) Temperature dependence of a superlattice peak at (6,1,0).  The solid line is a fit using the squared BCS function.  (c) $\theta$-2$\theta$ scan around (0,0,12) above (dashed line) and below (solid line) $T_{MI}$.  The scales of 2$\theta$ for the $T$ = 12 K and 80 K data are depicted at the bottom and top of the frame, respectively.}
\end{figure}
%=========================================================

The filled skutterudite compound PrRu$_4$P$_{12}$, which is a metal at room temperature, has attracted a lot of attention due to interest in the origin of the M-I transition at $T_{MI}$ = 60 K \cite{sekine97}.  Compared to other filled skutterudites, which show similar M-I transitions, PrRu$_4$P$_{12}$ is unique since it has neither a magnetic anomaly at $T_{MI}$, as seen in TbRu$_4$P$_{12}$ \cite{sekine2000}, GdRu$_4$P$_{12}$ \cite{sekine2000}, NdFe$_4$P$_{12}$ \cite{meisner} and SmRu$_4$P$_{12}$ \cite{sekine-Sm}, nor anti-quadrupolar ordering as observed in PrFe$_4$P$_{12}$ \cite{Hao}.  On the other hand, a subtle lattice distortion has been detected by electron diffraction measurements, which indicates weak superlattice spots at $(h,k,l) (h + k + l =$ odd) below $T_{MI}$ \cite{lee2001}.  To clarify the mechanism of the M-I transition, including the possibility of Fermi surface nesting, the precise crystal structure of the low temperature phase of PrRu$_4$P$_{12}$ must be identified.  To this end, we have carried out crystal structure analysis using electron and x-ray diffraction techniques.

Single crystals of PrRu$_4$P$_{12}$ were grown by a Sn flux method, which is described in detail elsewhere \cite{lee2001}.  For electron diffraction measurements, the as-grown crystals were thinned down to 50 $\mu$m by mechanical polishing and then ion milled using an argon ion-beam thinning apparatus.  For x-ray diffraction measurements, on the other hand, the as-grown crystals were mechanically polished into a nearly spherical shape with a diameter of ~250 $\mu$m.  The x-ray diffraction measurements were conducted using synchrotron radiation at BL-1B and BL-4C of the Photon Factory in Tsukuba, Japan.  The incident energy was fixed at 18 keV or 18.8 keV, close to the maximum energy where sufficient incident x-ray intensity could be obtained.  At BL-4C, a HUBER four-circle diffractometer was used with a scintillation counter as a detector.  At BL-1B, on the other hand, an imaging plate was used to record the Bragg spot intensity with rotating the single crystal around an axis perpendicular to incident x-ray beam (the $\theta$ axis) over a range of $5^{\circ}$ during each scan.  Electron diffraction measurements were carried out using a transmission electron microscope (JEOL 4000FX).  A convergent beam electron diffraction (CBED) technique was applied to study the space group of PrRu$_4$P$_{12}$ below $T_{MI}$, which is a powerful method for space group determination.  For cooling the samples down to 10 K, a double tilt liquid He holder was used.  The holder allows the samples to be tilted in a range of $\pm 15^{\circ}$ along two axes in the microscope.
%=========================================================
\begin{figure}
\includegraphics[width=\columnwidth]{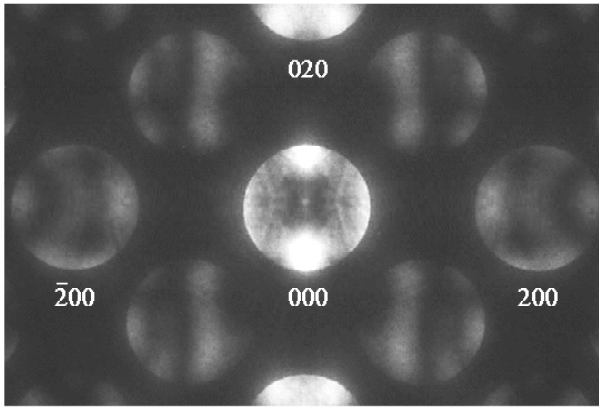}
\caption{\label{fig:CB-[001]} CBED pattern of the [0,0,1] zone-axis at $T$ = 10 K for PrRu$_4$P$_{12}$.  A 2-fold rotation symmetry is observed.}
\end{figure}
%=========================================================
%=========================================================
\begin{figure}
\includegraphics[width=\columnwidth]{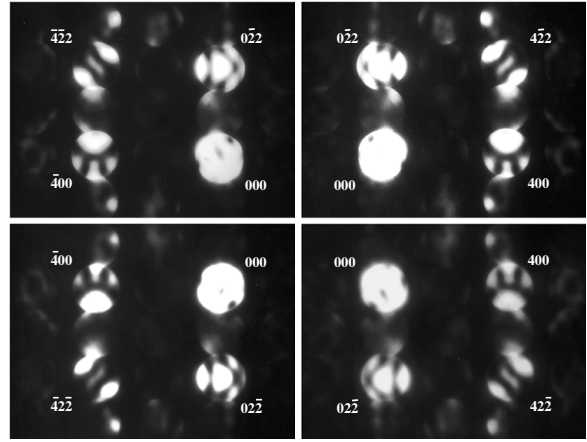}
\caption{\label{fig:CB-[101]} Dark-field CBED patterns of the [0,1,1] zone-axis at $T$ = 10 K for PrRu$_4$P$_{12}$.  The centers of each disk at (0,0,0), (0,2,$\bar{2}$), (4,0,0), (4,2,$\bar{2}$) and at counter positions with these reflections are located at exact Bragg positions.  Mirror symmetry is observed along the (1,0,0) plane.}
\end{figure}
%=========================================================

Figure 1(a) shows a superlattice peak of PrRu$_4$P$_{12}$ measured at low temperature by x-ray diffraction with an incident x-ray energy of 18.8 keV.  As shown, a well-defined peak appears below $T_{MI}$ at the commensurate position (0,0,11) consistent with ref.(8), indicating a structural phase transition from a body centred cubic (space group Im$\bar{3}$) to a lower symmetry structure.  The intensity of the superlattice peak decreases with increasing temperature and disappears at $T_{MI}$ (Figure 1(b)), suggesting that the corresponding lattice distortion relates to the M-I transition.  The solid line in Figure 1(b) depicts the result of a fit using the squared BCS function, which describes the temperature dependence of the peak intensity originating from a charge density wave (CDW) \cite{gruner}.  As shown, the solid line reproduces the observed data quite well, suggesting that a CDW could be present below $T_{MI}$.  To confirm whether the crystal structure maintains cubic symmetry below $T_{MI}$, profiles of the fundamental peaks were measured using a $\theta$-2$\theta$ scan.  If the crystal structure transforms from cubic symmetry, peak splitting is expected due to the formation of twin structures.  Figure 1(c) shows a comparison between the peak profiles at (0,0,12) for temperatures above and below $T_{MI}$.  As shown, the line-width of (0,0,12) is nearly constant, indicating that any possible peak splitting below $T_{MI}$ is smaller than $0.005^{\circ}$.  The lack of peak splitting has also been confirmed by powder x-ray diffraction measurements using synchrotron radiation.  These results suggest that the crystal structure below $T_{MI}$ is still cubic within an accuracy of 0.008 \% of the lattice constant.  This is in contrast with PrFe$_4$P$_{12}$, which transforms to an orthorhombic structure at low temperature \cite{iwasa}.

To identify the space group of the crystal structure below $T_{MI}$, CBED measurements were conducted.  Figure 2 shows a [0,0,1] zone-axis CBED pattern at $T$ = 10 K measured with an accelerating voltage of 100 kV.  In the figure, each disk shows a geometrical pattern due to multiple scattering, which reflects the symmetry of the crystal structure.  From the CBED pattern, it is clear that there is 2-fold rotation symmetry.  Combined with the fact that there is no extinction rule \cite{lee2001}, this observation suggests that the possible space group can be narrowed down to Pm$\bar{3}$ or P23.  These two space groups can be further distinguished by looking for the presence of mirror symmetry precisely along the (1,0,0) plane.
%=========================================================
\begin{table}
\includegraphics[width=\columnwidth]{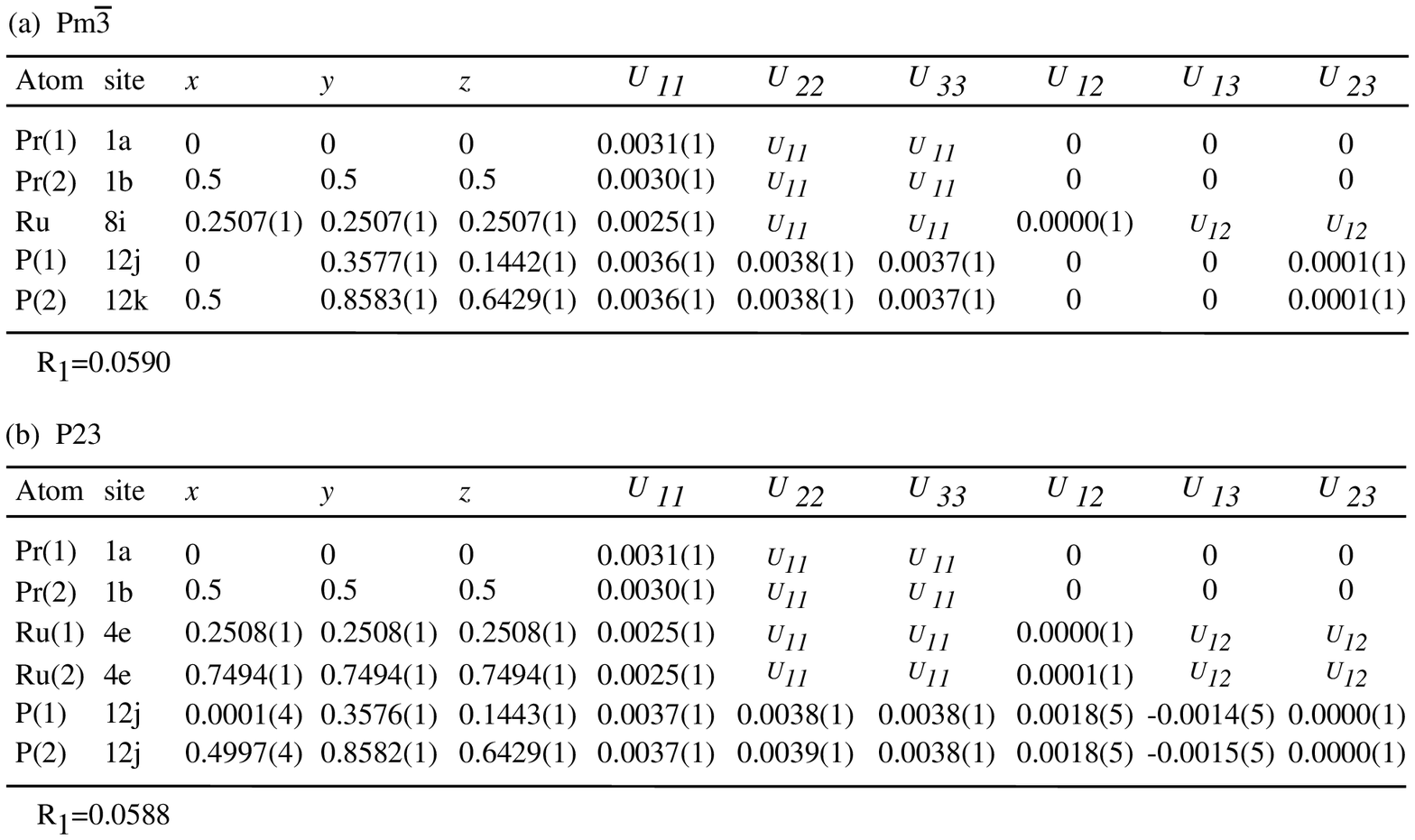}
\caption{\label{table:structure parameter} Crystal structure parameters of PrRu$_4$P$_{12}$ obtained from x-ray diffraction measurements at $T$ = 9 K, assuming space groups (a) Pm$\bar{3}$ and (b) P23.  $U_{ij}$ are the anisotropic thermal parameters, where the temperature factor is defined as exp[-2$\pi^{2}(h^{2}a^{*2}U_{11} + k^{2}b^{*2}U_{22} + l^{2}c^{*2}U_{33} + 2hka^{*}b^{*}U_{12} + 2hla^{*}c^{*}U_{13} + 2klb^{*}c^{*}U_{23}$)].  $a^{*}$, $b^{*}$ and $c^{*}$ are the inverse lattice parameters.  $R_1$ is an R-factor defined as $R_1$ = ($\Sigma ||F_o|-|F_c||)/\Sigma |F_o|$, where $F_o$ and $F_c$ are the observed structure factor and the calculated structure factor, respectively.  This table shows that the obtained parameters of Pm$\bar{3}$ and P23 are similar.}
\end{table}
%=========================================================

To examine for the presence of mirror symmetry more accurately, dark-field CBED measurements were conducted where the centers of the disks were located at exact Bragg positions.  Figure 3 shows [0,1,1] zone-axis dark-field CBED patterns over four quadrants measured with an accelerating voltage of 200 kV.  In each quadrant, the four disks are located at exact Bragg positions.  As shown, mirror symmetry along the (1,0,0) plane is present, suggesting that the space group is Pm$\bar{3}$.  The possibility of P23, however, still remains if the deviation from Pm$\bar{3}$ is very small.

To further refine the crystal structure, synchrotron x-ray diffraction measurements were conducted with an incident x-ray energy of 18 keV.  In these measurements, about 1200 independent Bragg spots including $\sim$600 superlattice spots were observed at $T$ = 9 K at a high S/N ratio using an imaging plate.  To analyze the results, the SHELX program \cite{shelx} was used after carrying out absorption corrections for spherical samples \cite{absor}.  Table 1 shows the obtained crystal parameters assuming the space group to be either Pm$\bar{3}$ or P23.  As shown, the atomic coordination for both space groups is the same within error, which suggests again that the higher symmetry group Pm$\bar{3}$ is reasonable for the insulator phase of PrRu$_4$P$_{12}$.

The obtained crystal structure of PrRu$_4$P$_{12}$ with the Pm$\bar{3}$ space group is depicted in Figure 4(a).  As shown, there are two Pr sites at the corners and the body center of the unit cell, surrounded by a P$_{12}$-icosahedron and a Ru$_8$-cube.  Compared to the Im$\bar{3}$ structure, the P(1) and P(2) atoms in the Pm$\bar{3}$ structure are displaced in a direction roughly perpendicular to the Pr-P bond.  In this distortion, the bond lengths of Pr(1)-P(1) and Pr(2)-P(2) are equivalent, 3.100(3) \AA\ at $T$ = 9 K.  On the other hand, the Ru atoms are displaced parallel to the Pr-Ru bond direction with changing the volume of Ru$_8$-cube, where the Pr(1)-Ru and Pr(2)-Ru bond lengths are 3.490(3) \AA\ and 3.470(3) \AA\ at $T$ = 9 K, respectively (Figures 4(b) and (c)).  As a consequence, a periodic ordering of two kinds of Ru$_8$-cube with different volumes appears.  In such a crystal structure, the electron density within the smaller Ru$_8$-cube around a Pr(2) atom can be higher than that around a Pr(1) atom, indicating the formation of a CDW.  Note that the valence of the Pr atom itself maintains a trivalent state independent of temperature as shown by Pr L$_2$-edge XANES measurements \cite{lee1999}.  This suggests that the CDW is produced by lattice distortions with each atom retaining its equilibrium charge.
%=========================================================
\begin{figure}
\includegraphics[width=\columnwidth]{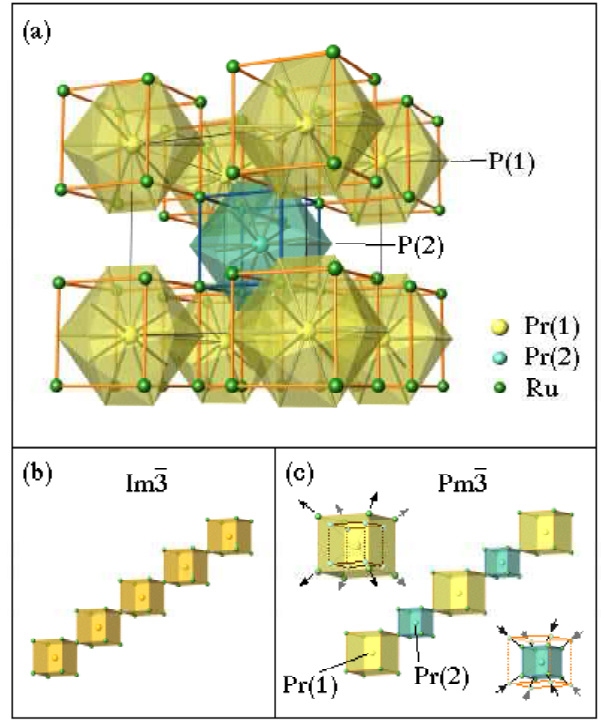}
\caption{\label{fig:crystal-structure} (a) Crystal structure of  PrRu$_4$P$_{12}$ in the Pm$\bar{3}$ phase.  (b, c) Schematic diagram of PrRu$_8$-cubes in (b) Im$\bar{3}$ and (c) Pm$\bar{3}$ phases connected at the corners in the (1,1,1) direction.  In the Pm$\bar{3}$ phase, two kinds of PrRu$_8$-cubes are present.  The displacement of the Ru atoms in the transition from the Im$\bar{3}$ to the Pm$\bar{3}$ phase is depicted in panel (c).}
\end{figure}
%=========================================================

To explain the M-I transition with CDW, one of key factor could be the nesting of the Fermi surface.  According to band calculations, the topology of the Fermi surface of PrRu$_4$P$_{12}$ in the Im$\bar{3}$ phase is nearly cubic with a volume of half the Brillouin zone, where a nesting instability is implied with wave vector of $\bf{q}$ = (1,0,0) \cite{harima2002}.  It has been suggested that the insulator phase of PrRu$_4$P$_{12}$ can appear in the crystal structure of the Pm$\bar{3}$ space group \cite{harima2003}.  Consistent with the prediction of the band calculations, the present work demonstrates that the crystal structure of PrRu$_4$P$_{12}$ takes the Pm$\bar{3}$ space group in the insulator phase.  Superlattice peaks are also observed below $T_{MI}$ at the same q-positions as expected by the nesting.  These agreements between the present observations and the band calculations suggest that the M-I transition in PrRu$_4$P$_{12}$ is attributed to a CDW phenomenon accompanied by perfect nesting of the Fermi surface.  Note that the perfect nesting phenomenon is rare in 3D systems like PrRu$_4$P$_{12}$ since three-dimensional Fermi surfaces are usually complicated.

There is, however, a counter-example with the above explanation.  For metallic LaRu$_4$P$_{12}$, although it has a similar Fermi surface to PrRu$_4$P$_{12}$, no M-I transition is observed \cite{sugawara,shirotani96,saha}.  This fact requires considering another mechanism.  Note that $f$ electrons are present in PrRu$_4$P$_{12}$ but not in LaRu$_4$P$_{12}$, it could be that $f$ electrons play an important role in the M-I transition.  For example, the $f$ electrons can assist the M-I transition by lowering their electronic energy below $T_{MI}$ with the development of the lattice distortion around Pr atoms.  It is also possible that both the $f$ electrons and the nesting phenomenon affect the M-I transition synergistically.  To clarify the mechanism of the M-I transition, it is valuable to further study the unique combination of $f$ electrons, lattice distortion and the nesting.

In conclusion, we have clarified by x-ray and electron diffraction techniques that the crystal structure of PrRu$_4$P$_{12}$ transforms from an Im$\bar{3}$ to a Pm$\bar{3}$ structure below $T_{MI}$, suggesting the formation of a CDW in the insulator phase.  The M-I transition could be attributed to the perfect nesting of the Fermi surface and/or to the instability of $f$ electrons.

The authors would like to thank K. Iwasa, M. Kohgi, K. Matsuhira and H. Harima for their helpful discussions.  This work was supported by a Grant-in-Aid for Scientific Research in Priority Area "Skutterudite" (No. 15072201) of the Ministry of Education, Culture, Sports, Science and Technology of Japan and a Grant from the Ministry of Economy, Trade and Industry of Japan.


\begin{thebibliography}{00}

\bibitem{sekine97} C. Sekine, T. Uchiumi, I. Shirotani and T. Yagi, Phys.
Rev. Lett. {\bf 79}, 3218 (1997).

\bibitem{shirotani97} I. Shirotani, T. Uchiumi, K. Ohno, C. Sekine, Y. Nakazawa, K. Kanoda, S. Todo and T. Yagi, 
Phys. Rev. B {\bf 56}, 7866 (1997).

\bibitem{meisner} G. P. Meisner, Physica B $\&$ C {\bf 108B}, 763 (1981).

\bibitem{bauer} E. D. Bauer, N. A. Frederick, P. -C. Ho, V. S. Zapf and M. B. Maple, 
Phys Rev. B {\bf 65}, 100506(R) (2002).

\bibitem{sales} B. C. Sales., D. Mandrus and R. K. Williams, Science {\bf 272}, 1325 (1996).

\bibitem{sugawara} H. Sugawara, Y. Abe, Y. Aoki, H. Sato, M. Hedo, R. Settai, Y. $\rm \bar{O}nuki$ and H. Harima, J. 
Phys. Soc. Jpn. {\bf 69}, 2938 (2000).

\bibitem{sekine2000} C. Sekine, T. Uchiumi, I. Shirotani, K. Matsuhira, T.Sakakibara,T. Goto and T. Yagi, 
Phys. Rev. B {\bf 62}, 11581 (2000).

\bibitem{sekine-Sm} C. Sekine et al., Science and Technology of High Pressure, 
ed. M. H. Manghnant et al., Universities Press, Hyderabad, India (2000) 826.

\bibitem{Hao} L. Hao, K. Iwasa, M. Nakajima, D. Kawana, K. Kuwahara, M.
Kohgi, H. Sugawara, T.D. Matsuda, Y. Aoki and H. Sato, Acta Physica Polonica B {\bf 34}, 1113 (2003).

\bibitem{lee2001} C. H. Lee, H. Matsuhata, A. Yamamoto, T. Ohta, H. Takazawa, K. Ueno, C. Sekine,
I. Shirotani and T. Hirayama, J. Phys.: Condens. Matter {\bf 13}, L45 (2001).

\bibitem{gruner} G. $\rm Gr\ddot{u}ner$, Rev. Mod. Phys. {\bf 60}, 1129 (1988).

\bibitem{iwasa} K. Iwasa, Y. Watanabe, K. Kuwahara, M. Kohgi, H. Sugawara, T. D. Matsuda, Y. Aoki and H. Sato, 
Physica B {\bf 312$\&$313}, 834 (2002).

\bibitem{shelx} G. M. Sheldrick, Acta Cryst. {\bf A46}, 467 (1990).

\bibitem{absor} {\it International Tables for Crystallography} (1995)
vol.C, edited by A. J. C. Wilson,
Kluwer Academic Publishers, Dordrecht, p. 520.

\bibitem{lee1999} C. H. Lee, H. Oyanagi, C. Sekine, I. Shirotani and M.
Ishii, Phys. Rev. B {\bf 60}, 13253 (1999).

\bibitem{harima2002} H. Harima, K. Takegahara, Physica B {\bf 312$\&$313}, 843 (2002).

\bibitem{harima2003} H. Harima, K. Takegahara, K. Ueda and S. H. Curnoe, Acta
Physica Polonica B {\bf 34}, 1189 (2003).

\bibitem{shirotani96} I. Shirotani, T. Adachi, K. Tachi, S. Todo, K. Nozawa, T. Yagi and M. Kinoshita, 
J. Phys. Chem. Solids {\bf 57}, 211 (1996).

\bibitem{saha} S. R. Saha, H. Sugawara, R. Sakai, Y. Aoki, H. Sato, Y. Inada, H. Shishido, R. Settai, Y. $\rm \bar{O}nuki$ and H. Harima, Physica B {\bf 328}, 68 (2003).


\end{thebibliography}
\end{document}